\documentclass[10pt,letterpaper]{article}
\usepackage[top=0.85in,left=2.75in,footskip=0.75in]{geometry}

\usepackage{changepage}

\usepackage[utf8]{inputenc}

\usepackage{textcomp,marvosym}

\usepackage{fixltx2e}

\usepackage{amsmath,amssymb}

\usepackage{cite}

\usepackage{nameref,hyperref}

\usepackage[right]{lineno}

\usepackage{microtype}
\DisableLigatures[f]{encoding = *, family = * }

\usepackage{rotating}

\raggedright
\usepackage[aboveskip=1pt,labelfont=bf,labelsep=period,justification=raggedright,singlelinecheck=off]{caption}

\makeatletter
\renewcommand{\@biblabel}[1]{\quad#1.}
\makeatother

\date{}

\usepackage{lastpage,fancyhdr,graphicx}
\usepackage{epstopdf}
\fancyheadoffset[L]{2.25in}
\fancyfootoffset[L]{2.25in}
\begin{document}
\vspace*{0.35in}

\begin{flushleft}
{\Large
\textbf\newline{Transient superdiffusion and long-range correlations in the motility patterns of trypanosomatid flagellate protozoa}
}
\newline
\\
Luiz~G.~A.~Alves\textsuperscript{1,2,3,*}, 
D\'ebora~B.~Scariot\textsuperscript{4}, 
Renato~R.~Guimar\~aes\textsuperscript{1,3}, 
Celso~V.~Nakamura\textsuperscript{4}, 
Renio~S.~Mendes\textsuperscript{1,3}, 
Haroldo~V.~Ribeiro\textsuperscript{1}
\\

\bigskip
\bf{1} Departamento de F\'isica, Universidade Estadual de Maring\'a, Maring\'a, PR, 87020-900,  Brazil
\\
\bf{2} Department of Chemical and Biological Engineering, Northwestern University, Evanston, IL, 60208, United States of America
\\
\bf{3} National Institute of Science and Technology for Complex Systems, CNPq, Rio de Janeiro, RJ, 22290-180, Brazil
\\
\bf{4} Departamento de Ci\^encias B\'asicas da Sa\'ude, Universidade Estadual de Maring\'a, Maring\'a, PR, 87020-900, Brazil 
\\
\bigskip

%
%





* lgaalves@dfi.uem.br

\end{flushleft}
\section*{Abstract}
We report on a diffusive analysis of the motion of flagellate protozoa species. These parasites are the etiological agents of neglected tropical diseases: leishmaniasis caused by \textit{Leishmania amazonensis} and \textit{Leishmania braziliensis}, African sleeping sickness caused by \textit{Trypanosoma brucei}, and Chagas disease caused by \textit{Trypanosoma cruzi}. By tracking the positions of these parasites and evaluating the variance related to the radial positions, we find that their motions are characterized by a short-time transient superdiffusive behavior. Also, the probability distributions of the radial positions are self-similar and can be approximated by a stretched Gaussian distribution. We further investigate the probability distributions of the radial velocities of individual trajectories. Among several candidates, we find that the generalized gamma distribution shows a good agreement with these distributions. The velocity time series have long-range correlations, displaying a strong persistent behavior (Hurst exponents close to one). The prevalence of ``universal'' patterns across all analyzed species  indicates that similar mechanisms may be ruling the motion of these parasites, despite their differences in morphological traits. In addition, further analysis of these patterns could become a useful tool for investigating the activity of new candidate drugs against these and others neglected tropical diseases.


\section*{Introduction}	
Kinetoplastids protists are responsible for numerous diseases in humans and animals. Many of these protozoa are the etiological agents of neglected tropical diseases~\cite{Nussbaum}. These diseases affect the lives of approximately one billion people around the world~\cite{Manderson} and are considered a serious public health problem in several countries. The main regions affected are developing countries located in tropical areas, where the parasites have appropriate natural conditions for life-cycle and insects vectors are abound~\cite{cdc}. Leishmaniasis, Chagas disease, and African sleeping sickness are examples of neglected tropical diseases caused by kinetoplastids parasites --- {\it Leishmania} spp, {\it Trypanosoma cruzi}, and {\it Trypanosoma brucei}, respectively~\cite{who}. These protozoa have a complex life-cycle, alternating between invertebrate (vector) and vertebrate hosts. These parasites have a flagellum at least during one of the evolutionary forms of their life-cycle~\cite{Milton,Kennedy,Grab}. The flagellum is a multifunctional organelle, responsible for cell propulsion and associated with control cell morphogenesis, chemotaxis, and cytokinesis process during last stage of the cell division cycle~\cite{Langousis}. The parasites motility is a key to host-cell attachment invasion and colonization of host tissues~\cite{Forestier,Hill,Ginger,Rodriguez,Langousis,Alizadehrad}.

Diffusive motion is ubiquitous in nature and plays a fundamental role in the motility of swimming microorganisms~\cite{Thurner,Metzler,Theves1,Theves2, Song, Amselem,Makarava}. Researchers in statistical mechanics have focused on phenomena where anomalous diffusion are present~\cite{Metzler}. Also, the field of non-extensive statistical mechanics has made an advance in the understanding of several systems where Boltzmann thermodynamic fails to explain the results~\cite{Tsallis,Plastino,Upadhyaya}. For instance, anomalous diffusion and non-Gaussian velocity distribution have been observed in the context of {\it Hydra} cells~\cite{Upadhyaya}, where maximum entropy densities associated with non-standard entropic measures were used to describe the motion of these cells. These densities are related to nonlinear diffusive process such as the generalized Fokker-Planck equation proposed in the context of the non-extensive statistical mechanics~\cite{Plastino}. Motility and diffusive patterns have also been investigated in protozoa~\cite{Rodriguez,Uppaluri,Heddergott,Forestier,Ballesteros,Hill,Broadhead,Zarley}. For instance, {\it T. brucei} studies have focused on the movement of propulsion~\cite{Rodriguez}, flexibility and directionality~\cite{Uppaluri}, and body adaptations to the environment~\cite{Heddergott}. Cell-host interaction~\cite{Forestier} for {\it Leishmania} spp and the flagellar beating~\cite{Ballesteros} for {\it T. cruzi} have also been studied. Motility is strongly related to cell viability in all flagellate kinetoplastid species and it is widely used as a proxy measurement for viability~\cite{Hill,Broadhead,Zarley}.

Each protozoan specie has unique adaptations depending on the different living conditions. The investigation of the dynamics, diffusion and motion behavior of these microorganisms is an advance in the understanding of the microbial pathogenesis mechanism and in the field of diffusive patterns. However, a more complete and general understanding of motility patterns of these parasites is still lacking. To overcome this gap, we study the diffusive dynamics of causative agents of the neglected tropical diseases. By tracking the positions of these parasites, we present a complete characterization of their motility patterns. Specifically, we show that the spread of the trajectories is superdiffusive for short-times and that probability distributions related to the radial positions differs from the predictions of the usual diffusion equation. We further verify that the velocity time series of individual trajectories have long-range correlations and are well approximated by a generalized gamma distribution. Our results reveal some ``universal'' parasite motility patterns that could facilitate the identification of novel targets for therapeutic intervention. Furthermore, it could be expanded to screen aspects of cell viability.

\section*{Materials and methods}
\subsection*{Parasites maintenance}
{\it Leishmania amazonensis} (MHOM/BR/75/Josefa strain) and {\it Leishmania braziliensis} promastigote forms were cultivated inside cell culture flasks containing Warren's medium (brain heart infusion plus bovine hemin and folic acid, pH 7.2) supplemented with 10\% of fetal bovine serum (FBS). The parasites were incubated at 25$^{\circ}\mathrm{C}$ for 48 h. {\it Trypanosoma cruzi} epimastigote forms (Y strain) were maintained in LIT medium (Liver Infusion Tryptose, pH 7.2), supplemented with 10\% of FBS and incubated at 28$^{\circ}\mathrm{C}$ during 96 h. Trypomastigote forms of {\it Trypanosoma brucei brucei} (EATRO-427 strain) were cultivated in HMI-9 medium, supplemented with 10\% of FBS, incubated at 37$^{\circ}\mathrm{C}$ and 5\% CO$_2$ tension for 24 h. These incubation periods are essential to harvest the protozoa in the  exponential growth phase.

\subsection*{Experimental setup, image acquisition and tracking}

After the  incubation periods, we have prepared a suspension containing about $6 \times 10^6$ parasites for {\it Leishmania} species, {\it  T. cruzi},  and {\it T. brucei} in Warren, LIT, and DMEM (Dulbecco's Modified Eagle Medium, supplemented with 2 mM L-glutamine, pH 7.4), respectively. The mediums were not supplemented with FBS. Next, 10 $\mu$L of the protozoa suspensions were placed between a glass slide and coverslip to start the image acquisition. The thickness between the glass slide and coverslip is comparable to the size of the protozoa ($\thicksim 5\,\,\mu$m)~\cite{Rozenberg,Farrar,deNoya}, reducing it to a two-dimensional problem. We used the Motic's BA410E microscope equipped with a 5.0 Megapixel CMOS camera at a resolution of $800 \times 600$ pixels, acquisition rate of 10 frames/second and magnification of $20 \times$. The area covered by the microscope at this configuration is $285.12 \times 213.84\,\,\mu$m$^2$. The length of the acquired videos was 10 minutes for each sample. We repeated this procedure three times for each protozoan. In order to extract the trajectories from the videos, we used a Matlab algorithm of motion tracking in image sequences~\cite{Wauthier}. After, we excluded the trajectories with less than 500 time-steps to ensure that we have long enough trajectories  for statistical analysis. For these trajectories, we removed the first and last 50 steps due to imprecision of the algorithm in track the protozoa positions. By visual inspection, we further removed the trajectories for which the algorithm mistook two or more microorganisms at some step along the path. All trajectories were smoothed by applying a moving average filter of length 10 and are available in \nameref{S1_dataset}. The viscosity of the culture medium was measured in a viscometer (Visco Star Plus) at 25 $^{\circ}\mathrm{C}$  and 50 rpm.  The viscosity values and the number of trajectories analyzed for each protozoan are shown in Table~\ref{tab1}. 

\begin{table}[!ht]

\caption{{\bf Dataset summary}}\label{tab1}
\begin{small}
\begin{tabular}{lclc}\hline
Protozoan & $\#$ Tracks & Medium & Viscosity [mPa/s]\\ \hline
\textit{L. amazonensis} & 105 & Warren & 0.38\\
\textit{L. braziliensis} & 135 & Warren & 0.38 \\
\textit{T. brucei} & 131 & DMEM & 0.71\\
\textit{T. cruzi} & 153 & LIT  & 0.39 \\ \hline
\end{tabular}
\end{small}

\end{table}

\begin{figure}[!h]
\begin{adjustwidth}{-2.25in}{0in}

\centering{
\includegraphics[scale=0.39]{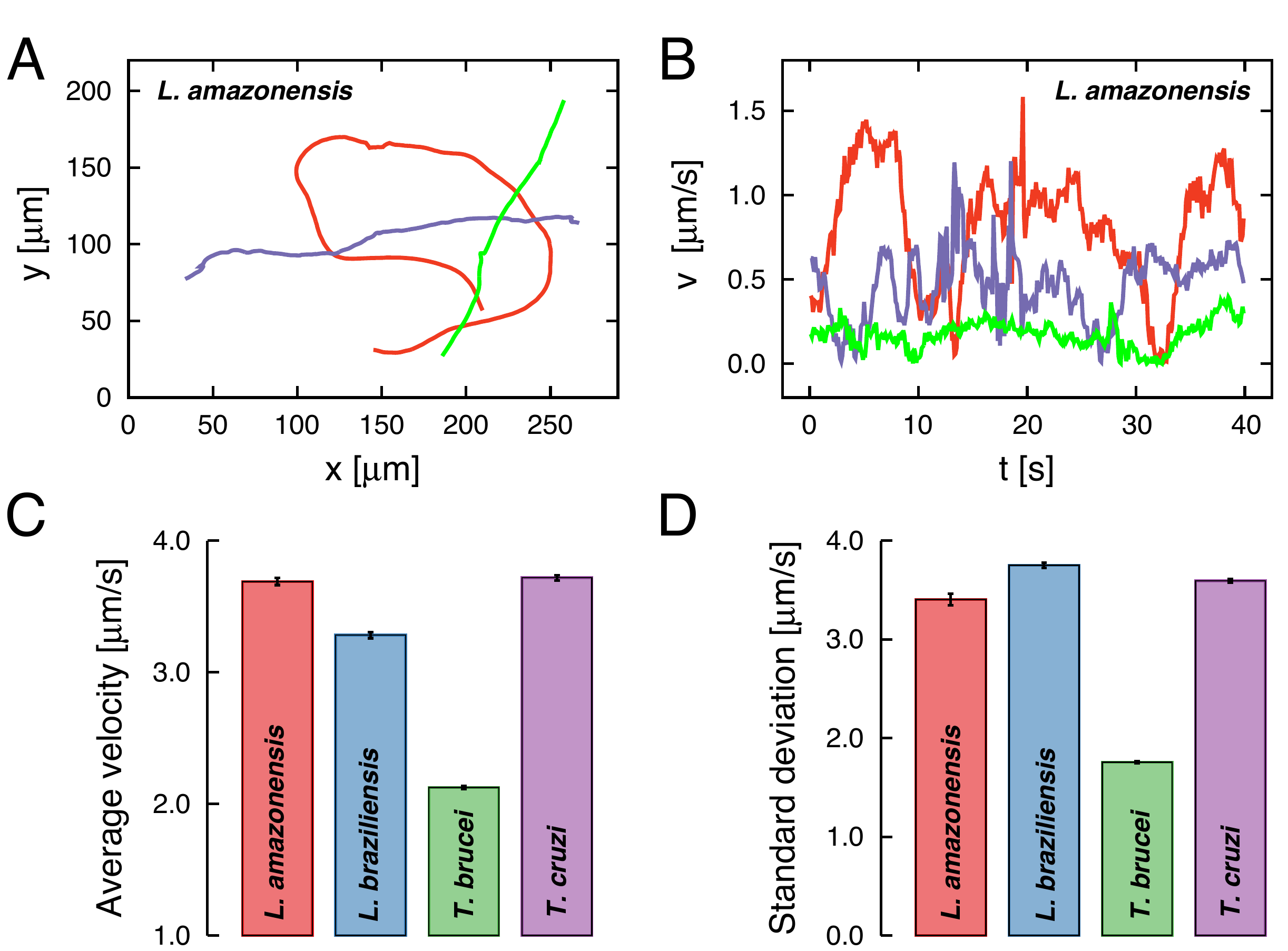}
}
\caption{ {\bf Illustration of trajectories and velocities}. A) Typical swimming trajectories of the \textit{L. amazonensis} and B)  the corresponding time series of the radial velocities $v(t)$ are shown. See Fig.~\nameref{S1_Fig} for other protozoa. C) Average velocities and D) standard deviation over all trajectories for each protozoan are represented in the bar plots. The error bars are 95\% confidence intervals calculated via bootstrapping.}
\label{fig:1}
\end{adjustwidth}
\end{figure}

In Fig.~\ref{fig:1}A, we show typical trajectories of the \textit{L. amazonensis}, that is, $\vec{r_i}(t)=[x_i(t),y_i(t)]$, where $x_i(t)$ and $y_i(t)$ are the horizontal and vertical components of the position vector $\vec{r_i}(t)$ in the time $t$ for the $i$-th track. In Fig.~\ref{fig:1}B, we plot the corresponding radial velocity time series, $v_i(t)=\sqrt{v_{x_i}^2(t)+v_{y_i}^2(t)}$, where $v_{x_i}(t)=[x_i(t+\Delta t)-x_i(t)]/\Delta t$ and $v_{y_i}(t)=[y_i(t+\Delta t)-y_i(t)]/\Delta t$ with $\Delta t=1/10$ s. Examples of trajectories and velocity time series of the other protozoa are shown in~\nameref{S1_Fig}. From these figures, we can observe circular patterns that could be related to possible hydrodynamic interactions with the walls, as previously observed in the motion of the bacteria {\it Escherichia coli}~\cite{Lauga}. Additional experiments considering different boundaries and thickness between the glass slide and the coverslip could further evaluate the effects of the walls on the motility of these protozoa. We have also calculated the average (Fig.~\ref{fig:1}C) and standard deviation (Fig.~\ref{fig:1}D) of the velocities for each protozoan specie. It is worth noting that the velocities of the swimming microorganisms depend on the viscosity of the culture medium~\cite{Heddergott} and other values can be found in the literature due to different viscosity~\cite{Heddergott, Rodriguez,Ballesteros, Uppaluri}. For instance, it was found that the wildtype bloodstream form of the \textit{T. brucei} can reach much higher velocities in the blood (around $30\,\mu$m/s)~\cite{Heddergott}.  {\it Trypanosoma} spp use the motility as a tool to evade the immune cells and remove the bind between their surface and antibodies molecules. In the blood vessels, where the protozoa is adapted to survive, red and white blood cells behave as support for the flagellum to propel the cell body. The same behavior is observed in more viscous liquids, justifying the increase in speed of the parasite in these environments~\cite{Heddergott}. Other factors that affect the velocities include chemical cues, oxygen content, pressure, flow and confinement~\cite{Heddergott}. Our results about the motility of {\it Leishmania} spp promastigote and {\it T. cruzi} epimastigotes suggest a very similar behavior profile.  Further details about these similarities are given in the results and discussion section.

\section*{Results and Discussion}

\begin{figure}[!ht]
\begin{adjustwidth}{-2.25in}{0in}
\centering{
\includegraphics[scale=0.39]{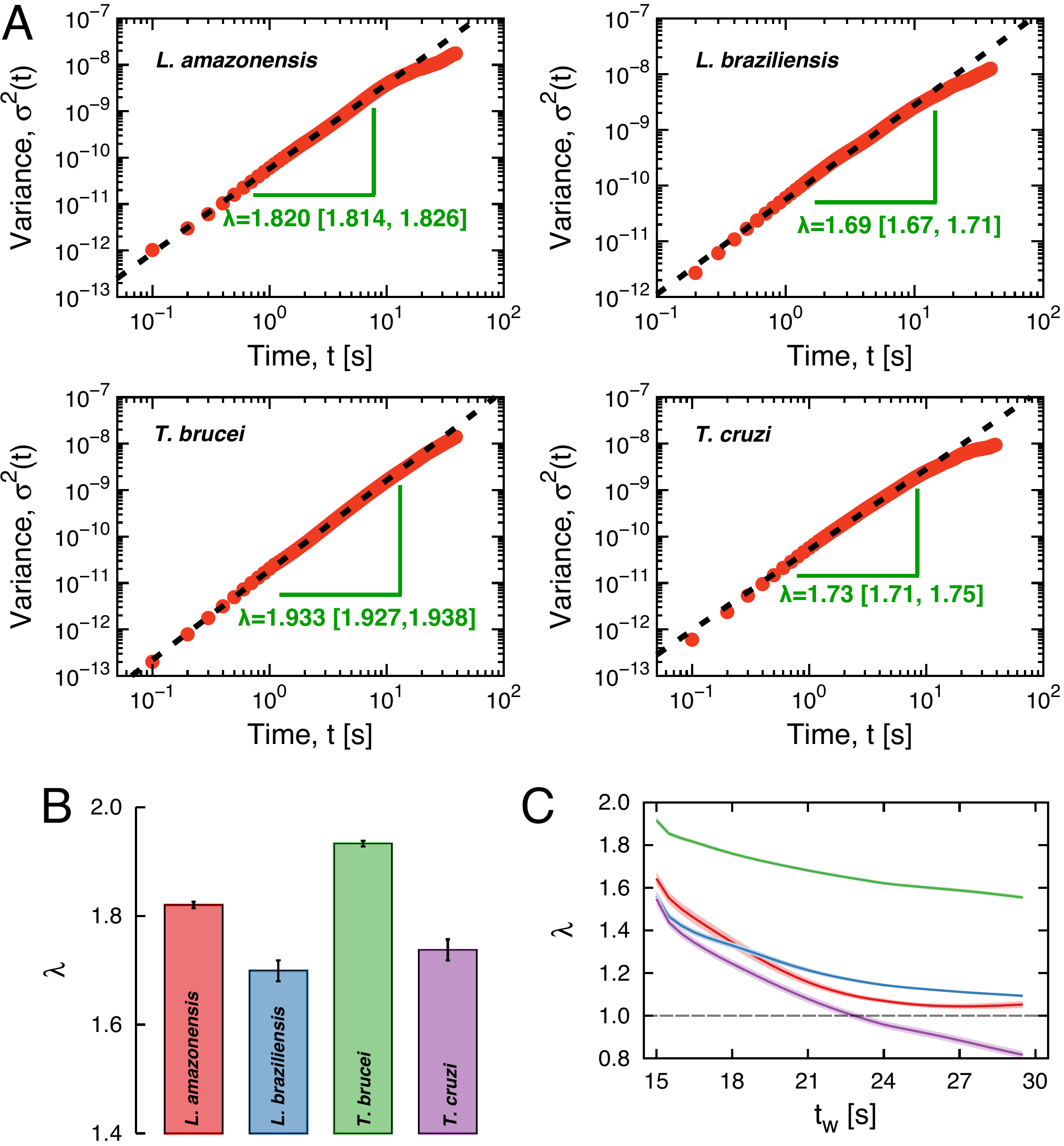}
}
\caption{{\bf The transient superdiffusive spreading of the protozoan trajectories}. A) Experimental values of the variance of the radial positions $\sigma^2(t)$ (red dots) for all species studied here (as indicated in the plots) are shown in log-log scale. The dashed lines represent power-law relationships $\sigma^2(t) \sim t^\lambda$, where the values of $\lambda$ were obtained by least square fitting a linear model to these log-log relationships (considering $t<10$ seconds). The values of $\lambda$ (and their 95\% bootstrap confidence intervals) are shown in plots, and in panel B) they are presented in bar plots, where error bars correspond to the 95\% bootstrap confidence intervals. C) Exponents $\lambda$ calculated within a window of size 30 seconds centralized in $t_w$ as a function of $t_w$. The different colors represent the four protozoa according to the ones used in Fig.~\ref{fig:2}B. The small shaded regions represent 95\% bootstrap confidence intervals and the dashed gray line represents the usual regime ($\lambda=1$).}
\label{fig:2}
\end{adjustwidth}

\end{figure}

We start by characterizing the spreading of the trajectories of all protozoa. In order to do so, we have considered the time series of the magnitude  of the position vector $\vec{r}_i(t)$ after subtracting its initial position $\vec{r}_i(0)=[x_i(0),y_i(0)]$. We thus measure the spreading by evaluating the time dependence of the variance of the centered radial position
\begin{eqnarray}\label{eq:msd}
\sigma^2(t) &=& \langle [\vec{r}_i(t)-\langle \vec{r}_i(t) \rangle]^2 \rangle \nonumber\\
&=&\frac{1}{N_k(t)-1}\sum_{i=1}^{N_k(t)} (\vec{r}_i(t)-\langle \vec{r}_i(t)\rangle)^2\,,
\end{eqnarray}
where $\langle \vec{r}_i(t) \rangle = \frac{1}{N_k(t)}\sum_{i=1}^{N_k(t)} \vec{r}_i(t)$ is the average radial position and $N_k(t)$ is the number of available trajectories for the $k$-th specie longer than $t$ steps (see Table~\ref{tab1}). For usual diffusive processes the variance is expected to { increase} linearly on time, that is, $\sigma^2(t) \sim t$. This usual behavior (or the Brownian motion) is related to absence of memory along the particle trajectory as well as indicates a finite characteristic scale for the position increments and for the waiting times between flights. However, more complex diffusive processes often display deviations of this linear behavior. When this happens, a typical behavior for the variance is a power-law dependence on time~\cite{Klafter}, 
\begin{equation}
\sigma^2(t) \sim t ^\lambda \,, 
\end{equation}
where $0<\lambda<1$ corresponds to  subdiffusion and $\lambda>1$ to  superdiffusion. In our case, the evolution of the variances are shown in Fig.~\ref{fig:2}A, where is evident that the spreading of the protozoan trajectories occurs much faster than the expected by a Brownian motion. We further observe an approximate power-law behavior over two decades of the temporal scale ($t<10$ seconds). By least square fitting the log-log relationships ($\log \sigma^2(t)$ versus $\log t$), we find that values of $\lambda$ are actually much larger than one. As shown in Fig.~\ref{fig:2}B, $\lambda$ ranges from $1.69$ for the \textit{L. braziliensis} to $1.93$ for the \textit{T. brucei}; therefore, the four flagellate protozoa studied here display a strongly superdiffusive behavior for short-times. A similar exponent was observed for the motion of intracellular particles with $\lambda\approx1.8$~\cite{Reverey}. It is worth note that, biological swimmers can present some persistence related to their drive mechanism for short-times. We expect that the diffusion may eventually approach the usual regime for longer trajectories. Thus, the values of $\lambda$ obtained here represent a initial short-time behavior, and other regimes could be observed for longer trajectories. In fact, we note from Fig.~\ref{fig:2}A that the curves start to bend downwards for $t>10$s. In Fig.~\ref{fig:2}C, we show the exponent $\lambda$ calculated within a window of size 30 seconds centralized in $t_w$ as a function of $t_w$, where we note that the values of $\lambda$ decrease with $t_w$ and approach the value expected by the usual diffusion.

Although all studied species showed superdiffusive spreading for short-times, we noticed a intriguing aspect of movement related to velocities and diffusion analysis for {\it T. brucei}. This protozoan shows the smallest velocity and the greatest diffusion exponent in culture medium, suggesting that the {\it T. brucei} motility is more directional than the other protozoa species considered in this work. Directional motility probably occurs because {\it T. brucei} is a free extracellular parasite and spreads across several tissues. Trypomastigote forms of {\it T. brucei} are highly adapted to live in intercellular spaces~\cite{Barry,Donelson,Lythgoe}. During spreading in the host {\it T. brucei} parasites penetrate between cells in the tissue where there are collagen fibers that can facilitate or hinder motion. For trypanosomes to reach the maximum forward velocity a specific density of cells is required. If the density of obstacles resembles collagen networks the protozoa swim backwards in order to avoid getting trapped~\cite{Heddergott}. These parasites have an entire antigen surface called variant surface glycoprotein (VSG)~\cite{Pays}. The VSG is different among individuals in a population of {\it T. brucei} and prevents specific binding with antibodies that could kill the parasites~\cite{Pays}. We hypothesize that the presence of surface molecules such as VSG's, which reduces the difficulty of the cell motion in its surroundings media, allows {\it T. brucei} parasite to diffuse faster.

Epimastigote forms of {\it T. cruzi} and promastigote forms of {\it Leishmania} spp exhibit similar velocities and diffusion exponent. These forms are faster than the {\it T. brucei} trypomastigote forms, have smaller diffusion exponent and have different strategies in this stage of the life-cycle. In the natural environment, {\it T. cruzi} epimastigote and {\it Leishmania} spp promastigote do not need to travel long distances. Specifically, epimastigote forms of {\it T. cruzi} must migrate to the midgut of the insect vector during a specific period of the life-cycle for proliferation~\cite{Souza}. In the case of promastigote forms of {\it Leishmania} spp, they must be phagocytized by mammalian cells to complete their life-cycle. At this stage, the {\it Leishmania} parasites secrete a substance called promastigote secretory gel (PSG). The PSG is a mucin-like gel and has been shown to be an important factor for the amastigote growth in the intracellular environment~\cite{Bates,Rogers}. PSG production seems to be responsible for the formation of agglomerates (rosettes) and could play a role as constraint for the diffusive motion~\cite{Rogers2,Rogers3}. Overall, we suggest that changes in motility parameters and surface molecules that affect mechanical or physical constraints restricting the ability of cells to freely diffuse could significantly contribute to the virulence of these parasites.

Another striking feature of diffusive processes is related to the probability distributions of the positions. For the usual diffusion with radial symmetry, this distribution can be obtained by solving the following differential equation
\begin{equation}\label{eq:diff}
\frac{\partial P(\vec{r},t)}{\partial t} = D \nabla^2 P(\vec{r},t)
\end{equation}
where $D$ is the diffusion coefficient and $\nabla^2$ is the two-dimensional Laplacian. By considering $P(r\to\infty,t)\to0$ and $P(\vec{r},0)=\delta^2(\vec{r}\,)$, we can show that probability distribution of the radial position is
\begin{equation}\label{eq:gaussnd}
P(r,t)=  \frac{r}{2\, D\, t}  \exp\left(\frac{-r^2}{4\, D\, t}\right)\,.
\end{equation}
It is worth noting that this solution leads to a linear behavior for the variance over time, that is, $\sigma^2(t) = 4 D t$. Furthermore, the distributions given by the Eq.~\ref{eq:gaussnd} are self-similar in time and collapse into a single curve, 
\begin{equation}\label{eq:sgusual}
P(\xi)= 2\, \xi  \exp\left(-\xi^2\right)\,,
\end{equation}
for the rescaled position $\xi(t)=r(t)/\sigma(t)$. The usual diffusion equation (Eq.~\ref{eq:diff}) and its solution (Eq.~\ref{eq:gaussnd}) can be understood as a null model for the distributions of the radial positions of the protozoa if we assume that they behave as Brownian particles, and deviations from this prediction is another indication of anomalous diffusion.

\begin{figure}[!h]
\begin{adjustwidth}{-2.25in}{0in}
\centering{
\includegraphics[scale=0.39]{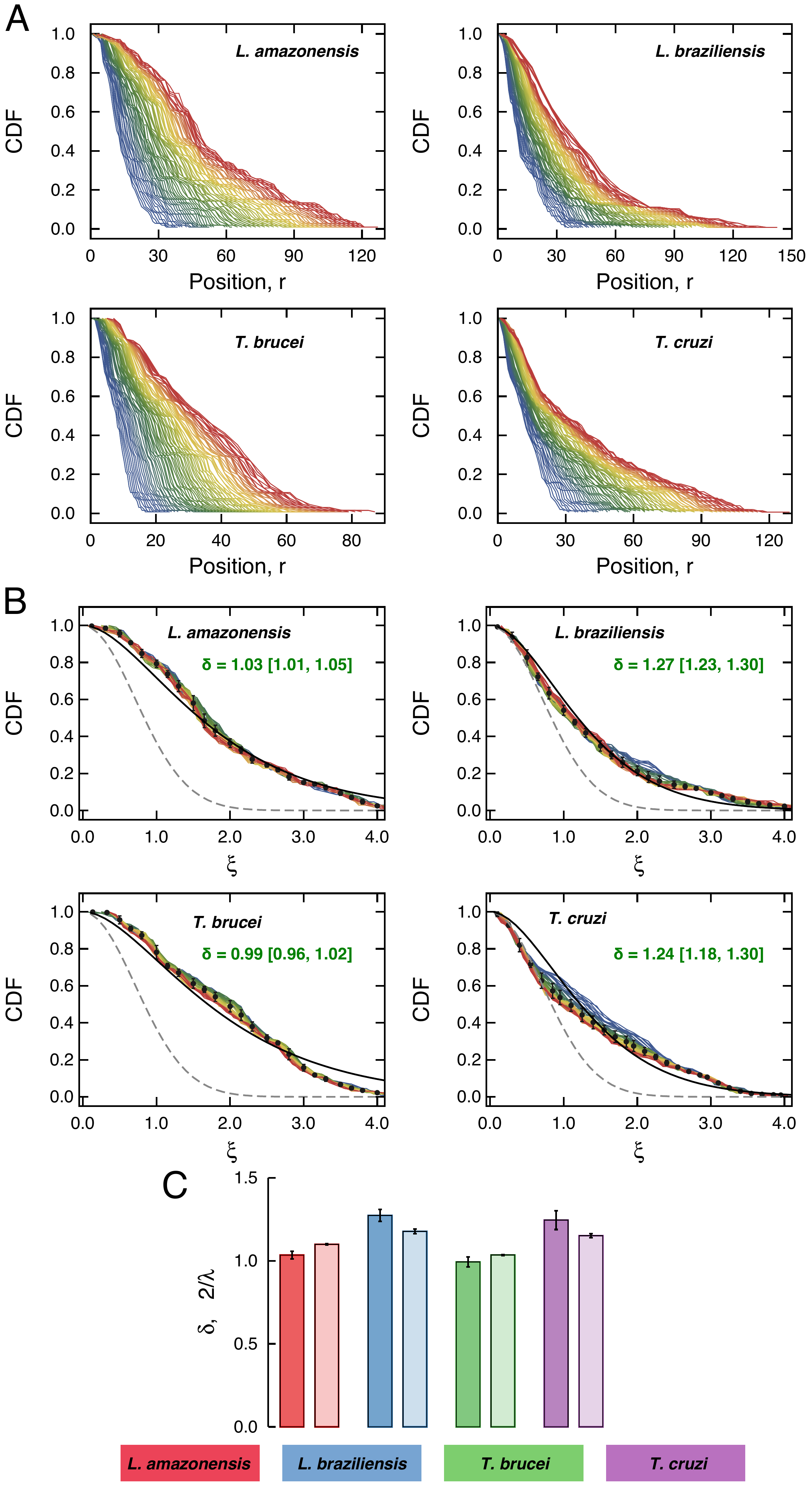}
}
\caption{{\bf The self-similar nature of the protozoan positions}. A) Time evolution of the cumulative distribution functions (CDF) of the radial position $r(t)$ for each protozoan. The color code indicates the time $t$, ranging from blue ($t=2$s) to red ($t=10$s). B) The same distributions after considering the rescaled position $\xi(t)= r(t)/\sigma(t)$. The black dots represent window averages of the CDFs and the error bars stand for one standard deviation. The gray dashed lines represent the distribution given by the normalized Gaussian distribution (Eq.~\ref{eq:sgusual}). The continuous lines are the cumulative version of the distribution of Eq.~\ref{eq:sgc} and the best values for the fitting parameter $\delta$ are shown in the plots. C) Comparison between the values of $\delta$ obtained via least squares fitting the window averages of the CDFs and the ones expected by the time dependence of the variances, that is, $\delta=2/\lambda$.
}\label{fig:3}

\end{adjustwidth}

\end{figure}

We have calculated the time evolution of the empirical distributions of the radial positions. As shown in Fig. \ref{fig:3}A, we note that these distributions shift toward positive values of $r$ while also become broaden over time $t$. After considering the rescaled position $\xi(t)=r(t)/\sigma(t)$, we observe that all distributions collapse into a single curve (Fig.~\ref{fig:3}B). This result demonstrates the empirical self-similar nature of the protozoa trajectories, but also reveals remarkable deviations between the empirical distributions and the expected by the usual diffusion equation (gray dashed lines in Fig.~\ref{fig:3}B). 

In an attempt to find a better description for these empirical distributions, we propose to replace the Gaussian term in Eq.~\ref{eq:gaussnd} by a stretched Gaussian (characterized by another parameter $\delta>0$). Specifically, we have considered the following probability distribution for the radial positions
\begin{eqnarray}\label{eq:sg}
P(r,t)= \frac{ \delta}{\Gamma(2/\delta)} \frac{r}{(2\,D\, t)^{2/\delta}}  \exp\left(\frac{-|r|^\delta}{2\, D\, t}\right)\,,
\end{eqnarray}
where $\Gamma(x)=\int_0^\infty y^{x-1}e^{-y}dy$ is the gamma function. We observe that $\delta=2$ recovers the usual Gaussian term; however, for $0<\delta<2$ the tail of this distribution goes to zero slower than the usual case, whereas for $\delta>2$ it decays faster. A similar generalization was proposed by Richardson~\cite{Richardson} in the context of atmospheric diffusion by considering a spatial-dependent diffusion coefficient, and is considered one of the first anomalous diffusion equation. The distribution given by the Eq.~\ref{eq:sg} is also self-similar and for the rescaled radial position $\xi(t)=r(t)/\sigma(t)$ it can be written as
\begin{equation}\label{eq:sgc}
P(\xi)=\frac{\delta}{\Gamma\left(2/\delta\right)}\,\xi\,e^{-\left| \xi\right| ^{\delta }}\,.
\end{equation}
We have thus adjusted Eq.~\ref{eq:sgc} to the window average values of the empirical distributions via least squares fitting. The continuous lines in Fig.~\ref{fig:3}B show that agreement is far from being perfect, but this generalization is a better description when compared with  the distribution emerging from the usual diffusion equation (Eq.~\ref{eq:sgusual}). Therefore, the diffusive motion of the protozoa studied here displays simultaneously an anomalous (enhanced) scaling of variance $\sigma^2(t)$ as well as radial position distributions with much longer tails than the expected by Brownian swimmers. The values of the best fitting parameters $\delta$ are shown in Fig.~\ref{fig:3}C, where we observe that the \textit{L. amazonensis} and the \textit{T. brucei} display larger tails ($\delta\approx 1$) when compared with the \textit{L. braziliensis} and the \textit{T. cruzi} ($\delta\approx 1.25$). It is also worth noting that the values of $\delta$ should be related to values of $\lambda$. In fact, the time dependence of the variance evaluated from Eq.~\ref{eq:sg} is $\sigma^2(t) \propto  t^{2/\delta}$, leading to $\delta=2/\lambda$, a relationship that is roughly valid for the values $\delta$ obtained via the fitting procedure (Fig.~\ref{fig:3}C). 

In order to further characterize the diffusive motion of the protozoa, we investigate the radial velocity time series $v_i(t)$ related to the individual trajectories (Fig.~\ref{fig:1}B). We first ask whether these time series have short or long-range correlations. To answer this question, we have applied the detrended fluctuation analysis (DFA)~\cite{Peng,Kantelhardt} to these time series. This technique consists of four steps: $i)$ first, we define the integrated profile $Y(t)=\sum_{k=1}^{t}[ v_i(k) - \bar{v}_i$], where $\bar{v}_i = 1/t_{\text{max}} \sum_t^{t_{\text{max}}} v_i(t)$ with $t_{\text{max}}$ being the length of the $i$-th time series; $ii)$ next, we split $Y(t)$ into $N_n=t_{\text{max}}/n$ non-overlapping segments of size $n$; $iii)$ for each segment, a local polynomial trend (here we have used a linear function, but higher orders do not change our results) is calculated and subtracted from $Y(t)$, defining $Y_n(t)=Y(t)-p_\nu(t)$, where $p_\nu(t)$ represents the local trend in the $\nu$-th segment; $iv)$ finally, we calculate the root-mean-square fluctuation function $F(n)=[\frac{1}{N_n}\sum_{\nu=1}^{N_n} \langle Y_n(t)^2\rangle_\nu]^{1/2}\,,$ where $\langle Y_n(t)^2\rangle_\nu$ is the mean square value of $Y_n(t)$ over the data in the $\nu$-th segment. If the velocity time series is self-similar, the fluctuation function $F(n)$ presents a power-law dependence on the time scale $n$,  that is, $F(n)\sim n^{h}$, where $h$ is the Hurst exponent. If $h=1/2$ the velocities are either uncorrelated or short-range correlated, whereas $h\neq1/2$ indicates that the time series is longe-range correlated. 

\begin{figure}[!h]
\begin{adjustwidth}{-2.25in}{0in}
\centering{
\includegraphics[scale=0.39]{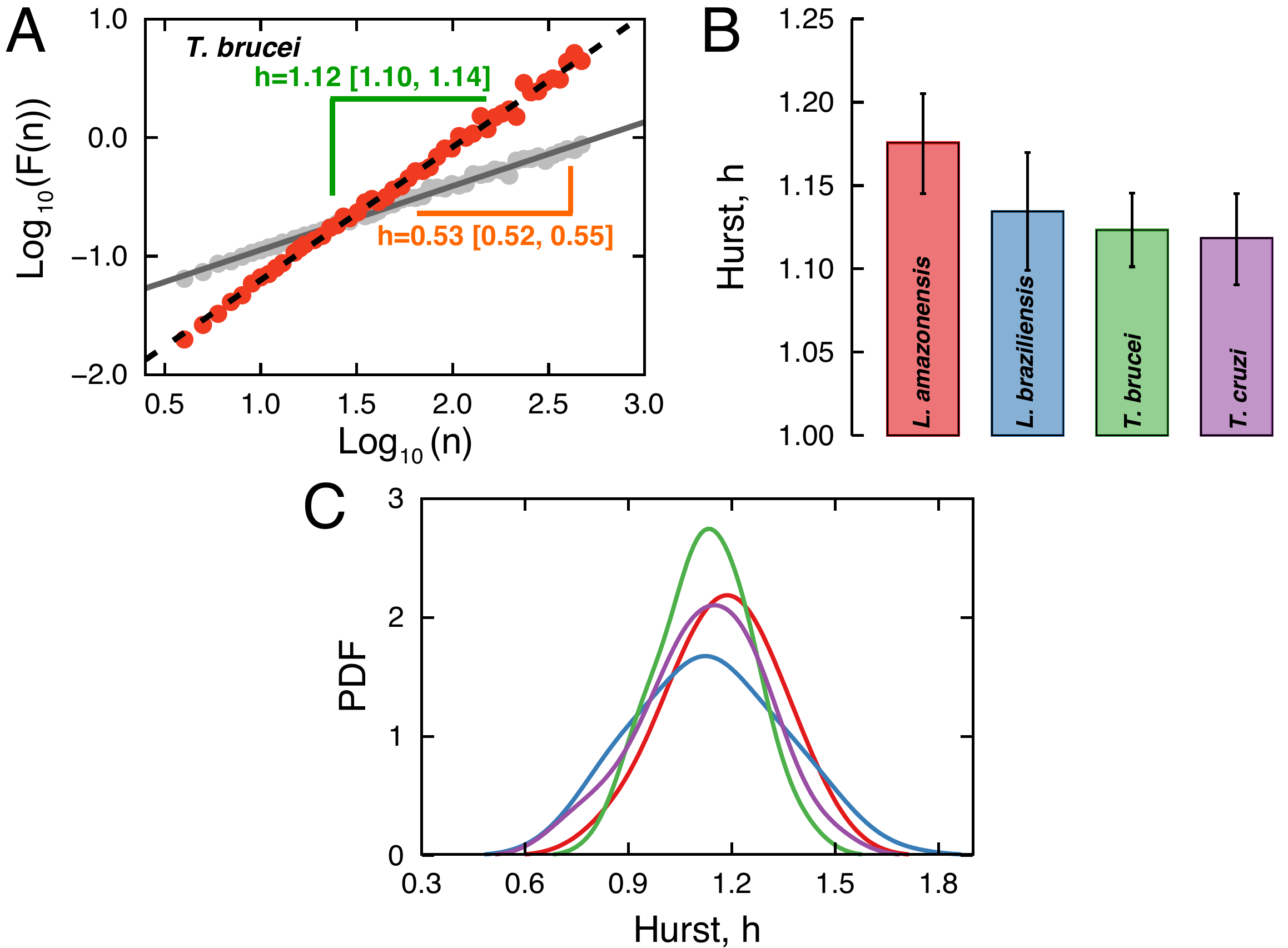}
}
\caption{{\bf Long-range and persistent correlations in the protozoan velocities}.  A) Detrended fluctuation analysis (DFA) of the velocity time series $v(t)$. A typical behavior for the fluctuation function $F(n)$ versus the scale $n$ (in a log-log plot) for a velocity time series of the protozoan \textit{T. brucei} (See~\nameref{S2_Fig} for other examples). The red dots are the values of $F(n)$ obtained for the original time series, while the gray ones were obtained for a random shuffled version of this time series. The black dashed line is a least squares fit to the relationship $\log_{10}[F(n)]$ versus $\log_{10}[n]$ for the original time series and the gray continuous line is the same for the random shuffled version. The value of Hurst exponent is $h=1.12$ for the original data and $h\approx0.5$ for the shuffled data (a similar behavior is observed for all time series). B) Average values of the Hurst exponent over all trajectories for each protozoan. The error bars are 95\% confidence intervals calculated via bootstrapping. C) Probability distribution function (PDF) of the Hurst exponents $h$ for each protozoan obtained via kernel density estimation method.}
\label{fig:4}
\end{adjustwidth}
\end{figure}

We have applied the above procedure to all velocity time series and a typical behavior for the fluctuation function $F(n)$ is depicted in Fig.~\ref{fig:4}A. In this log-log plot, we adjust a linear model for obtaining the Hurst exponent $h$, which is $h=1.12$ for the original time series and is close to $1/2$ for random shuffled versions of the time series. We calcule the Hurst exponent for all time series and the average values for each protozoan is shown in Fig.~\ref{fig:4}B. These averages are practically indistinguishable from each other and indicate that the velocities of the protozoa have long-range correlations. Furthermore, these velocities display a strong persistent behavior (since $h\approx 1$), that is, positive increments in the velocities are followed by positive increments and negative increments are followed by negative increments much more frequently than by chance. We have also evaluated the distributions of $h$ for each protozoan specie for all individuals (Fig.~\ref{fig:4}C), where we note that these distributions peak around $h\approx 1$ and that they are quite overlapped.

\begin{figure}[!h]
\begin{adjustwidth}{-2.25in}{0in}
\centering{
\includegraphics[scale=0.39]{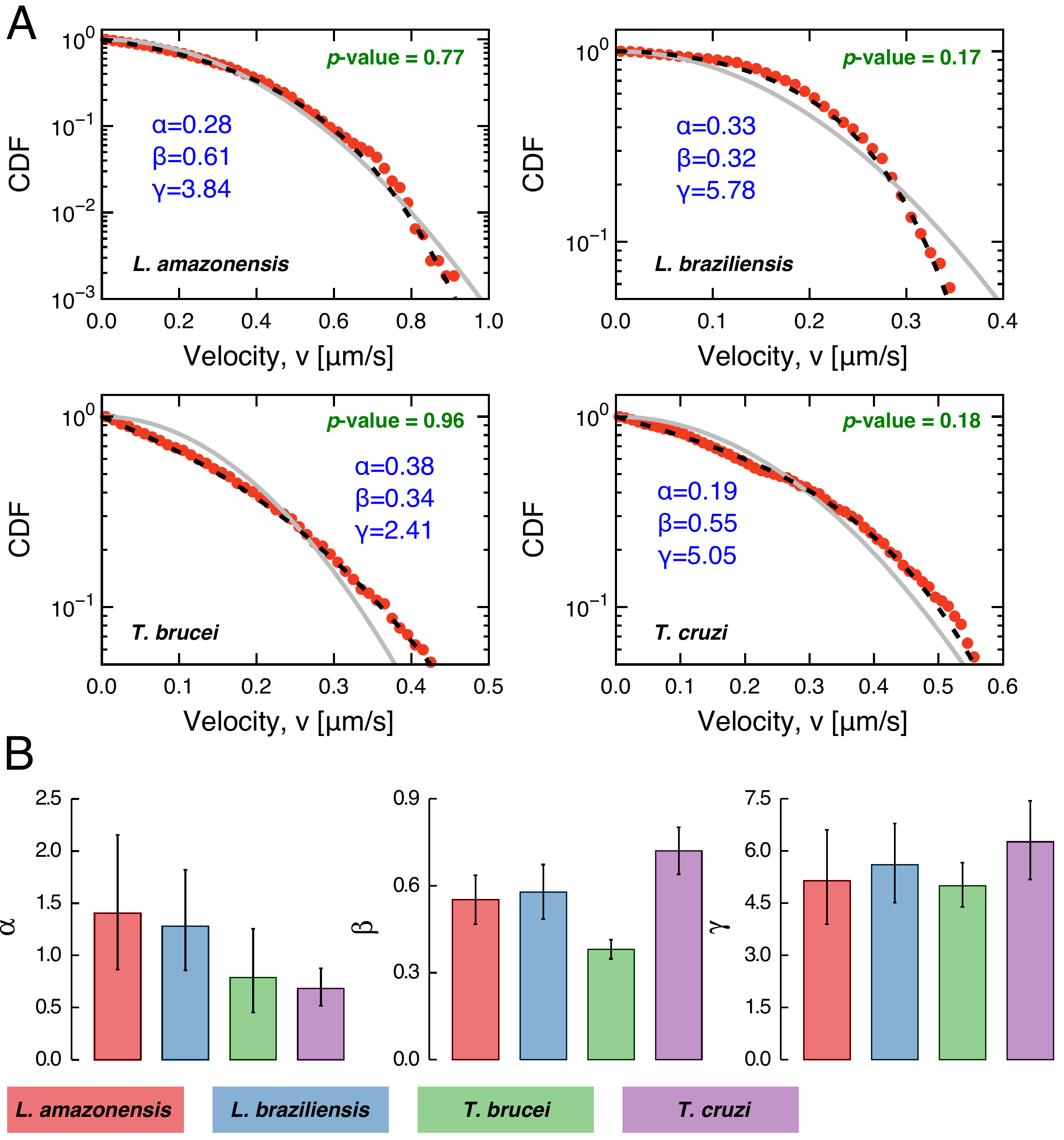}

}

\caption{ {\bf The velocities are not described by a two-dimensional Maxwell-Boltzmann distribution}. A) Typical examples of cumulative distribution functions (CDF) of the velocities $v(t)$ from a single trajectory of the protozoa. The gray continuous lines are the best fits obtained for the two-dimensional Maxwell-Boltzmann distribution (Eq.~\ref{eq:mb}), whereas the dashed black lines are the fits for the generalized gamma distribution (Eq.~\ref{eq:gamma}). The values of parameters $\alpha$, $\beta$ and $\gamma$ were obtained via maximum-likelihood method and are shown in the plots. We also present the $p$-values of the Kolmogorov-Smirnov test showing that we cannot reject the gamma hypothesis for these particular trajectories. We tested all trajectories and the KS test cannot reject this hypothesis for about 50\% of the trajectories (all $p$-values are shown in~\nameref{S3_Fig}). B) Average values of the best fitting parameters over all trajectories. The error bars are 95\% confidence intervals calculated via bootstrapping.}
\label{fig:5}
\end{adjustwidth}
\end{figure}

Another intriguing question is whether the velocity distributions of the protozoa exhibit a particular functional form (Fig.~\ref{fig:5}A). In this case, the two-dimensional Maxwell-Boltzmann distribution (or Rayleigh distribution)
\begin{equation}\label{eq:mb}
P(v)=\frac{2 v}{T} e^{-v^2/T}\,\qquad(v>0),
\end{equation}
is a natural null model for the protozoan velocity ($T$ is a parameter). This distribution represents the velocity of two-dimensional gas particles in thermodynamic equilibrium (at a temperature $T$) and also emerges when evaluating the distribution of the magnitude of velocity vectors whose the components are uncorrelated and normally distributed (with zero mean)~\cite{Atkins}. This functional form has been found to describe quite well the velocities of humans in a very peculiar situation (a \textit{mosh pit})~\cite{Silverberg} and it has also been used in the attempt of modeling the velocity distributions of {\it Hydra} cells in a two-dimensional setup~\cite{Upadhyaya}. Deviations from this model give clues about weather the velocities are correlated or not. In our case, we have tested the (two-dimensional) Maxwell-Boltzmann hypothesis by adjusting this distribution for each velocity time series and verifying the goodness of the fit via Kolmogorov-Smirnov test. This hypothesis was rejected for almost all velocity time series ($\approx99\%$), a result that somehow agrees with the more complex behavior observed in the correlation analysis of these time series. Aiming to find a better description for these empirical distributions, we have considered the generalized gamma distribution
\begin{eqnarray}
P({v})=\frac{\gamma}{\beta\,\,  \Gamma (\alpha )}\,  \left(v/\beta\right)^{\alpha\,  \gamma -1}\,e^{-\left(v/\beta\right)^{\gamma }},\,\qquad(v>0)
\label{eq:gamma}
\end{eqnarray}
where $\alpha$ and $\gamma$ are the shape parameters and $\beta$ is a scale parameter. Despite of being an \textit{ad hoc} generalization, it is worth noting that this distribution recovers the two-dimensional Maxwell-Boltzmann (for $\alpha=1$, $\gamma=2$ and $\beta^2=T$), and it has been employed by several authors as a wind speed model~\cite{Morgan,Kiss,Carta,Auwera}. For our data, the Kolmogorov-Smirnov test cannot reject the generalized gamma hypothesis in about 50\% of trajectories (see~\nameref{S3_Fig}), an improved description when compared with the two-dimensional Maxwell-Boltzmann distribution. Furthermore, we have tested for the usual gamma (Eq.~\ref{eq:gamma} with $\gamma=1$), log-normal, Weibull and $q$-exponential distributions, finding that they do not outperform the generalized gamma description. In Fig.~\ref{fig:5}B, we show the average values of the best fitting parameters $\alpha$, $\gamma$ and $\beta$. These values are practically indistinguishable among the four protozoa. An exception occurs for the \textit{T. brucei}, which is characterized by a significantly smaller value of $\beta$. This fact is a direct consequence of the small standard deviation observed for the velocities of \textit{T. brucei} (see Fig.~\ref{fig:1}D), since the standard deviation of $v$ calculated from Eq.~\ref{eq:gamma} is proportional to $\beta$.

\section*{Conclusions}
We presented a description of the motility patterns of four trypanosomatid flagellate protozoa. By analyzing the time evolution of the positions of these protozoa, we identified that the spreading of their trajectories are strongly superdiffusive for short-times and characterized by self-similar probability distributions with longer tails than the expected by the usual Brownian motion. We also investigated the velocities of these protozoa, finding out that they have long-range correlations and present a strong persistent behavior. We further observed that the velocity distributions cannot be described by a two-dimensional Maxwell-Boltzmann distribution (the natural candidate for a random process). Instead, a generalized gamma distribution showed to be a better description. Thus, our results show that the motility patterns of these protozoa are anomalous in several ways and also reveal that these four protozoa exhibit similar behaviors, despite their morphological differences. Deviations from the behaviors reported here can be employed as an indicator of drug activity in drug tests.

\section*{Authors contribution}
Conceived and designed the experiments: DBS, LGAA and RSM. Performed the experiments: DBS and LGAA. Analyzed the data: HVR, LGAA, RRG and RSM. Contributed reagents/materials/analysis tools: CVN, DBS, HVR, LGAA, RRG and RSM. Wrote the paper: CVN, DBS, HVR, LGAA, RRG and RSM. Prepared the figures: LGAA.

\section*{Funding}
This work has been supported by the agencies Conselho Nacional de Desenvolvimento Cient\'ifico e Tecnol\'ogico (CNPq), Coordena\c{c}\~ao de Aperfei\c{c}oamento de Pessoal de N\'ivel Superior (CAPES), and Funda\c{c}\~ao Arauc\'aria (FA). HVR thanks the financial support of CNPq (grant 440650/2014-3), LGAA thanks for the financial support of CAPES (grant 99999.006842/2015-01) and RSM thanks FA (grant 263/2014) for partial financial support. The funders had no role in study design, data collection and analysis, decision to publish, or preparation of the manuscript. 

\newpage 
\section*{Supporting Information}

\renewcommand\thefigure{S\arabic{figure}}    

\setcounter{figure}{0}    
\subsection*{S1 Dataset}
\label{S1_dataset}
{\bf Dataset containing the trajectories employed in this study.} The xls file contains one sheet for each protozoan specie. In each sheet, the columns represent the trajectories and the lines represent the time (in $1/10$ seconds scale). The cell value is the position $\{x,\,y\}$ for a given trajectory and time. 	

\subsection*{S1 Fig.}
\label{S1_Fig}
\begin{figure}[h!]
\begin{adjustwidth}{-2.25in}{0in}
\centering{
\includegraphics[scale=0.45]{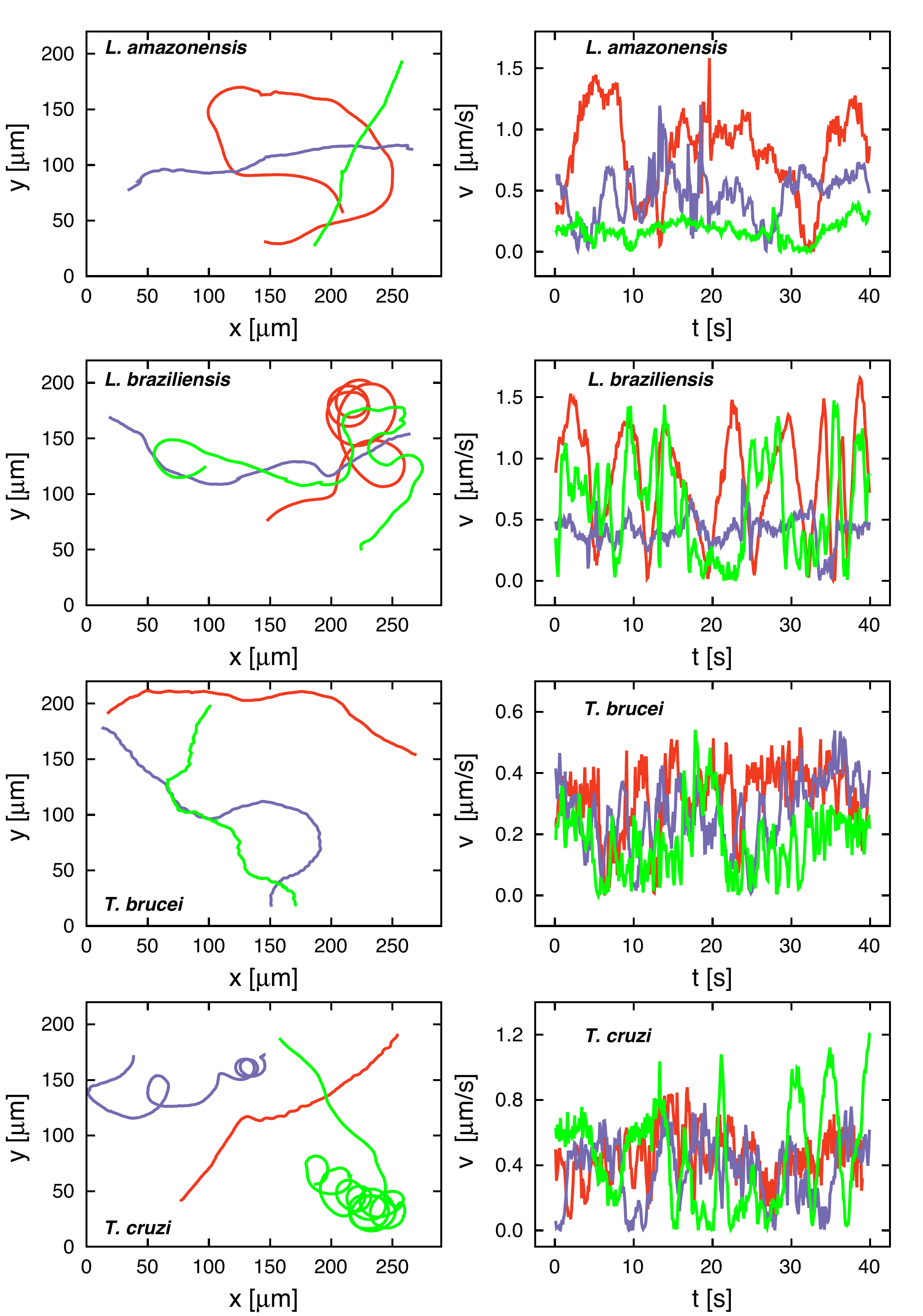}

}
\caption{{\bf Trajectories and velocities}. The left panel shows typical trajectories of the four protozoa. The right panel shows the corresponding velocity time series $v(t)$.}
\end{adjustwidth}
\end{figure}

\subsection*{S2 Fig.}
\label{S2_Fig}

\begin{figure}[h!]
\begin{adjustwidth}{-2.25in}{0in}
\centering{
\includegraphics[scale=0.45]{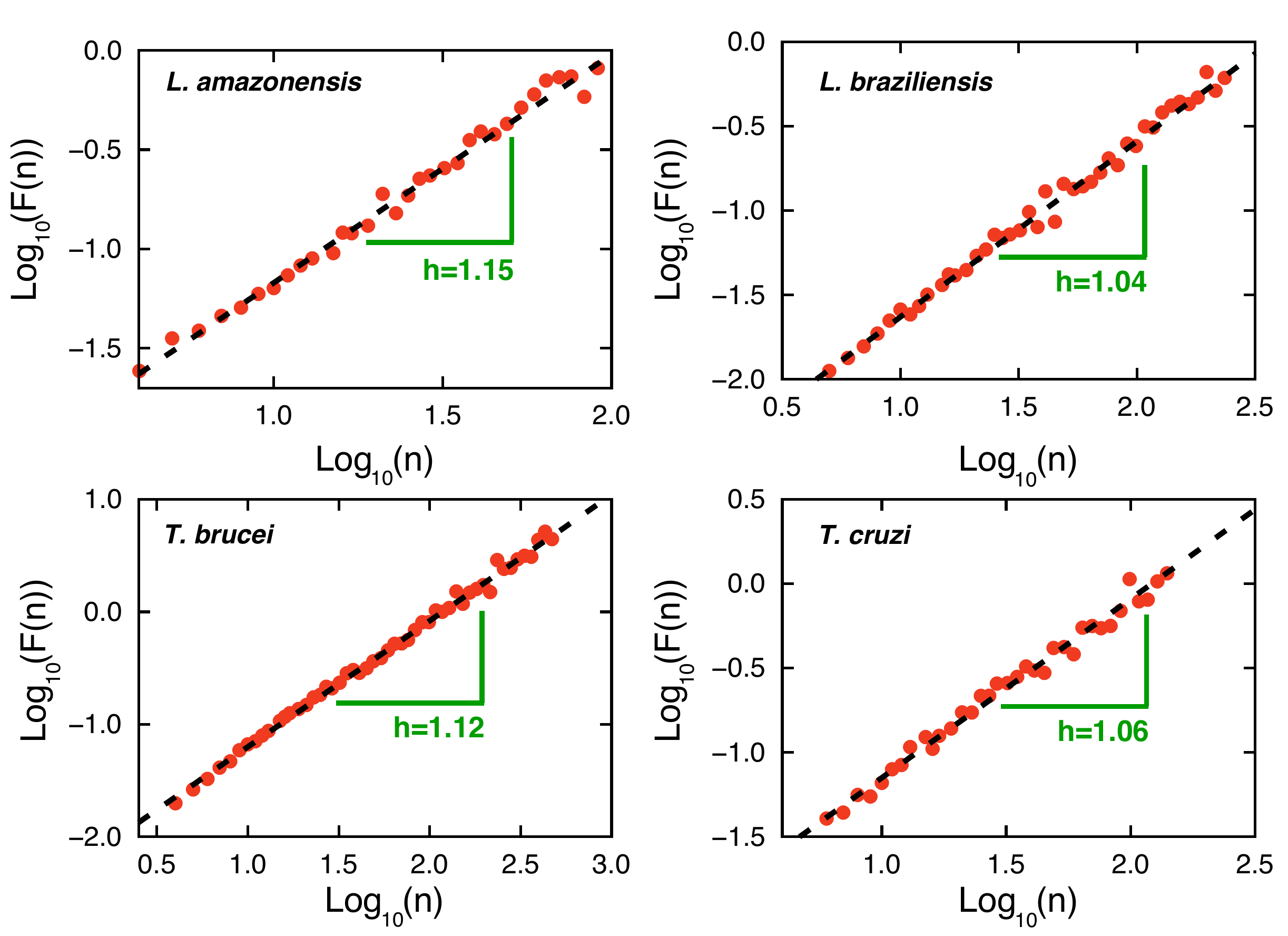}

}
\caption{{\bf Fluctuation function}. Another typical examples of detrended fluctuation analysis (DFA) for the velocity time series.}
\end{adjustwidth}
\end{figure}

\subsection*{S3 Fig.}
\label{S3_Fig}
\begin{figure}[h!]
\begin{adjustwidth}{-2.25in}{0in}
\centering{
\includegraphics[scale=0.45]{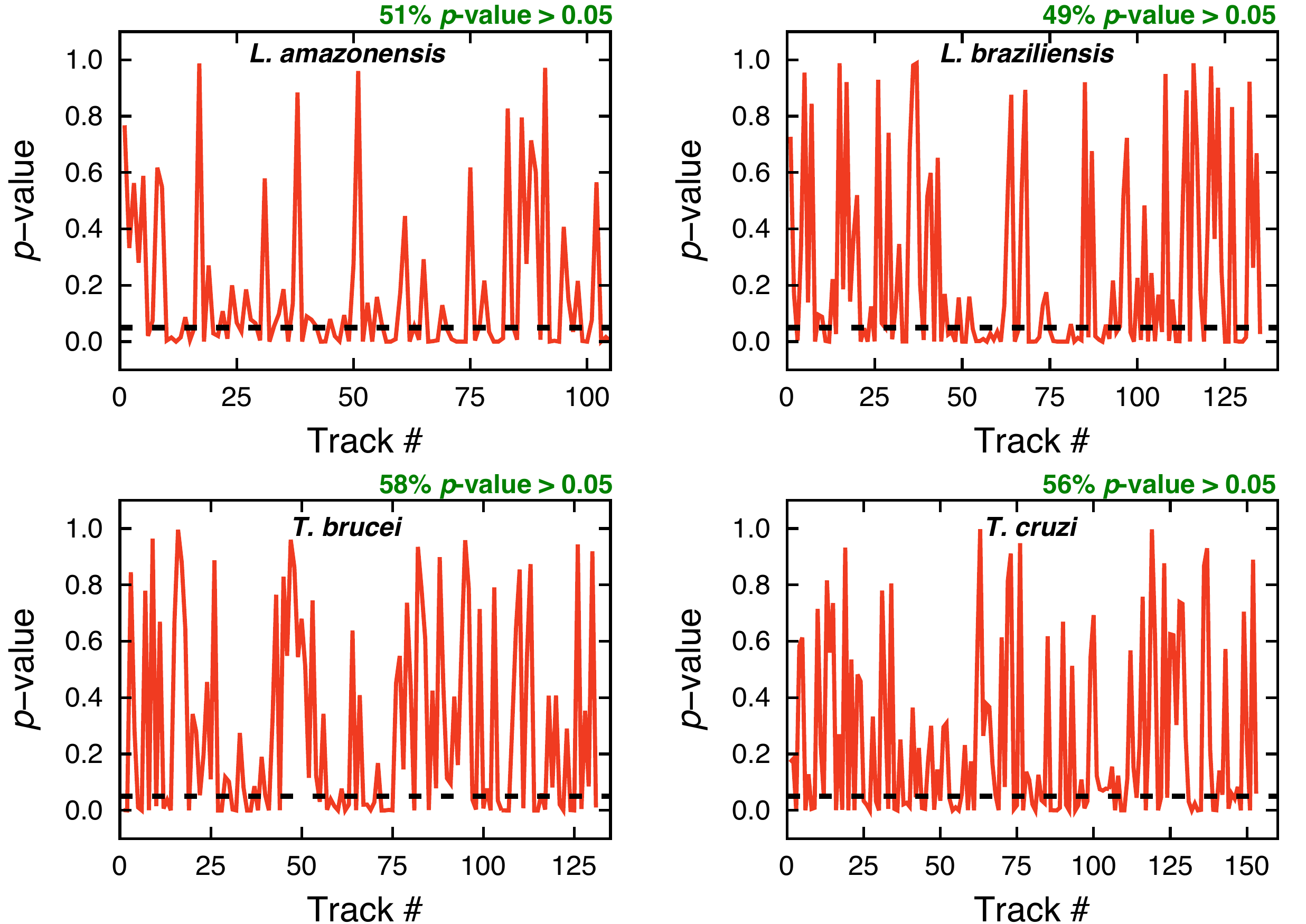}

}
\caption{{\bf Kolmogorov-Smirnov test}. The red lines show the $p$-values of Kolmogorov-Smirnov test of the generalized gamma hypothesis for all velocities time series. About 50\% of the velocity time series have $p>0.05$ (horizontal line).}
\end{adjustwidth}
\end{figure}

\section*{Acknowledgments}
We thank Peter B. Winter, Meagan Bechel, and Nicol\'as Pel\'aez for helping with the manuscript revision. 
\nolinenumbers

%
%
%

\end{document}